\documentclass{article}
\pdfoutput=1
\usepackage[utf8]{inputenc} 
\usepackage[T1]{fontenc}
\usepackage{float}
\usepackage{authblk}
\usepackage[normalem]{ulem}
\usepackage{color, colortbl}
\usepackage{amsmath}
\usepackage{amsthm}
\usepackage{adjustbox}
\usepackage{graphicx}
\usepackage{mathtools}
\usepackage{fullpage}
\usepackage{todonotes}
\usepackage{subcaption}
\usepackage[mathlines]{lineno}

\definecolor{gragreblu}{rgb}{0.04, 0.49, 0.65}
\usepackage[colorlinks = true,
            linkcolor = gragreblu,
            urlcolor  = gragreblu,
            citecolor = gragreblu]{hyperref}
            
\usepackage[style=numeric,sorting=none,]{biblatex}
\newcolumntype{P}[1]{>{\centering\arraybackslash}p{#1}}

\title{\textbf{Networks of monetary flow at native resolution}}
\author[1]{Carolina Mattsson\thanks{mattsson.c@northeastern.edu}}
\affil[1]{Network Science Institute, Northeastern University, Boston, MA, USA}

\begin{document}
\maketitle
\pagenumbering{arabic}

\begin{abstract}
People and companies move money with every financial transaction they make. We aim to understand how such activity gives rise to large-scale patterns of monetary flow. In this work, we trace the movement of e-money through the accounts of a mobile money system using the provider's own transaction records. The resulting transaction sequences---balance-respecting trajectories---are data objects that represent observed monetary flows. Common sequential motifs correspond to known use-cases of mobile money: digital payments, digital transfers, and money storage. We find that each activity creates a distinct network structure within the system, and we uncover coordinated gaming of the mobile money provider's commission schedule. Moreover, we find that e-money passes through the system in anywhere from minutes to months. This pronounced heterogeneity, even within the same use-case, can inform the modeling of turnover in money supply. Our methodology relates economic activity at the transaction level to large-scale patterns of monetary flow, broadening the scope of empirical study about the network and temporal structure of the economy.
\end{abstract}

\section*{Introduction}

The movement of money within an economy is primarily studied in aggregated form, using data on monetary flows between industries. The movement of money at smaller scales has long been impractical to consider empirically, and thus also under-explored conceptually. Modern payment infrastructure, however, is relatively centralized and increasingly digital. As people and companies conduct business, they are leaving a treasure trove of data about the real economy---at the finest possible resolution---on the servers of financial institutions worldwide. A small but growing group of researchers has begun to use such datasets to explore the economic and financial behavior of individuals~\cite{farrell_weathering_2015,blumenstock_airtime_2016,economides_mobile_2017,aladangady_transactions_2019}. Others have analyzed the financial transactions taking place within payment systems as networks, seeking to capture the overall structure of such systems~\cite{soramaki_topology_2007,kyriakopoulos_network_2009,bech_illiquidity_2012,kondor_rich_2014,zanin_topology_2016,iosifidis_cyclic_2018}. However, we currently lack a way to relate individual behavior to the structure of the system as is conveyed by the movement of money. 

We would like to be able to study how millions of individual transactions come together to create large-scale patterns in the movement of money. In this work, we address a concrete version of this question: how do we build a network representation of monetary flow from the financial transaction records of a payment system? Such a representation would encode the structure of monetary flow at the scale of the system, with a level of resolution equal to that at which money changes hands.

We consider a large dataset of mobile money transaction records from a provider in East Africa, which covers ten months of activity for millions of users. Mobile money is a new global industry that has expanded rapidly across Africa, South Asia, and Southeast Asia since the late 2000s~\cite{gsma_mobile_money_state_2015}. Mobile money providers support a digital version of the local currency (e-money). They host e-money accounts, process transfers, and service payments for users over their cellular infrastructure, where digital transactions are instantaneous. Digital services are facilitated by a large cadre of on-the-ground \textit{mobile money agents}. These agents represent the provider and are physically located in the area they service. Mobile money agents offer conversion between cash and e-money, as would a teller, but they run their own operations often in conjunction with a retail shop. Mobile money agents are paid on commission.~\cite{cull_agent_2018}

Well-known use cases for mobile money, such as digital payments, digital transfers, and money storage, generally involve several sequential transactions of different types. For instance, paying a bill using the mobile money system might entail first depositing cash and then making a digital payment. Typical sequential patterns, which we call motifs, are suitable for isolating the most common actions taken by mobile money users. To study these sequences empirically, we trace e-money as it moves through the mobile money system.

This paper defines a data transformation that turns financial transaction records into a dataset of observed transaction sequences---\textit{balance-respecting trajectories}. Each trajectory represents a specific amount of money observed to move through a specific sequence of accounts following a particular motif. In the language of monetary economics, balance-respecting trajectories represent observed monetary flows. We rely on the rules of basic accounting to build out trajectories, respecting the balance in every account at every point in time. Accounting guarantees that the movement of money is a conservative process; money does indeed \textit{flow}. In the language of the language of network science, conservative processes are \textit{walk processes} and balance-respecting trajectories are observed instances of a walk process.~\cite{pfitzner_betweenness_2013,rosvall_memory_2014,saramaki_exploring_2015}  

We trace e-money from when it enters the mobile money system to when it exits, and group observed trajectories by the motif they follow. We create aggregated \textit{entry-exit networks} where the nodes are the mobile money agents or corporations at the start and end of the observed trajectories. The links can be weighted to represent the movement of money, or the absolute flow of money, through the mobile money system as a whole. We focus on the trajectories that begin with cash deposits to agents, and the aggregated networks that gives each deposit equal weight. 

We discover that each user activity moves money through a different network structure at the system scale: digital payments result in a hub-and-spoke network, digital transfers form a largely amorphous network, while money storage and other activity that involves no digital transactions creates a network with geographic assortativity. Within this last network, we also uncover systematic gaming of the commissions system by a small subset of mobile money agents. This fraudulent behavior appears to be coordinated within scores of small, isolated groups of agents. In each case, trajectories let us observe individuals' actions and aggregate their effect on the movement of money up to the scale of the entire system. 

We also find that user activity moves e-money through the corresponding motifs in anywhere from minutes to months, thus returning that e-money to provider-facing accounts at substantially different rates. There are differences in these distributions between activities: commission gaming and bill payments happen considerably faster, on average, than do person-to-person transfers. But more importantly, the underlying distribution in return time for each of the activities range across several orders of magnitude. Empirical heterogeneity in turnover times greatly complicates estimation and interpretation of the velocity of money, a related theoretical concept from macroeconomics, at smaller scales. 

The methodology presented here brings the tools of network science, current and future, into reach for studying how money moves within payment systems. Conceptually, the network structure of money flow within an economy is a different angle from which to consider the interaction of scales in economics. Trajectory-based network analysis of empirical monetary flows at native resolution can quantify this structure. Network analysis could provide another way to measure the economic power of ``hubs'' (ex. large firms) or the economic independence of ``communities'' (ex. regions). Moreover, payment systems themselves are what connect the monetary system to the underlying economy. Balance-respecting trajectories remain interpretable as the flow of money, and can provide an empirical grounding for ambitious lines of inquiry in monetary economics. 

\section*{Mobile Money Data}

We analyze a mobile money dataset containing over 300 million transaction records generated by over 5 million anonymous users. This activity was facilitated by over 40,000 anonymous mobile money agents. Each record includes the sender, recipient, time stamp, amount, fee, type, and resulting balances of each transaction. The most common transaction types are summarized in Table \ref{table:txn_types}. Users can deposit money by giving cash to a mobile money agent, who then places e-money onto their account (cash-in). A withdrawal reverses this process (cash-out). Users can transfer e-money to other users using the person-to-person (p2p) service. Bill payment transactions (bill-pay) are payments to utilities or other large corporations. Mobile airtime (top-up) and mobile data (data) purchases are payments to the provider. Generally, mobile airtime and mobile data purchases are micro-transactions in that they are orders of magnitude smaller than other transaction types. 

\begin{table}[h]
    \begin{tabular}{|l|l|r|r|r|r|r|}
    \cline{1-7}
    \textbf{Type} & \textbf{Description} & \multicolumn{1}{l|}{\textbf{Records}} & \multicolumn{1}{l|}{\textbf{Amount}} & \multicolumn{1}{l|}{\textbf{Popularity}} & \multicolumn{2}{l|}{\textbf{Transactions}} \\ 
     & & & \textbf{Average} & & \textbf{Average} & \textbf{Median} \\ \cline{1-7}
    Cash-in  & Deposit via agent         & 24.1\% & \$42.96 & 95.9\% & 17.0 &  7 \\ \cline{1-7}
    Cash-out & Withdrawal via agent      & 19.3\% & \$47.91 & 94.3\% & 13.9 &  7\\ \cline{1-7}
    P2P      & Person-to-person transfer &  5.9\% & \$52.80 & 78.5\% &  5.1 &  4\\ \cline{1-7}
    Top-up   & Mobile airtime purchase   & 41.8\% & \$ 0.82 & 79.9\% & 35.4 & 13\\ \cline{1-7}
    Data     & Mobile data purchase      &  4.0\% & \$ 0.86 & 17.9\% & 15.1 &  4\\ \cline{1-7}
    Bill-pay & Bill payment              &  3.4\% & \$20.58 & 24.0\% &  9.5 &  3\\ \cline{1-7}
    \end{tabular}
    \caption{\textbf{Mobile money transaction types} This six most common transaction types in the mobile money data. The number of records of one type is reported as a percentage of all transaction records, and the average amount of these transactions is denoted in US Dollars at purchasing power parity (PPP) with the local currency. Popularity is the percentage of users who made at least one transaction of this type. The average or median number of transactions refers to those made by these users, excluding those with zero transactions of this type.}
    \label{table:txn_types}
\end{table}

Table \ref{table:txn_types} shows that most transactions are ones where e-money either enters or exits the system; the \textit{network boundary} is very prominent. Deposits and withdrawals of e-money are the most popular, in that such transactions are made by the largest fraction of users. Indeed, mobile money recipients often choose to withdraw their e-money into cash straight away rather than to keep it in their accounts or send it onward.~\cite{stuart_cash_2011}  Mobile airtime purchases are the most common type of transaction in the data, and they are indeed small. Person-to-person transfers, which keep e-money in circulation, are also popular but users make fewer of them so they are less common in the data. It is worth noting that surveys of users show that person-to-person transfers are the most popular service~\cite{intermedia_financial_2016} indicating that users do not consider deposits and withdrawals to be separate actions, necessarily.

Given the salience of the network boundary, we reconsider oft-described use cases for mobile money as sequential patterns of transaction types---motifs. Paying a bill using the mobile money system would generally entail a cash-in transaction followed by a bill-pay transaction. Similarly, e-money from a cash-in can be used to purchase mobile airtime or mobile data. These are all well-known use cases of mobile money systems~\cite{gsma_mobile_money_state_2015}. Another prototypical sequential pattern is the digital transfer motif, which involves three transactions: a cash-in, then a p2p, and then a cash-out~\cite{mbiti_home_2013}. Note that p2p transactions that are \textit{not} subsequently withdrawn keep money in circulation within the mobile money system. 

E-money from a cash deposit can also be withdrawn again into cash without undergoing any digital transactions at all. This creates an in-out motif that is fairly common in mobile money systems, and there are several accepted explanations. Economides \& Jeziorski (2017) describe this use case as money storage, a way to avoid carrying cash while travelling and to avoid storing cash at home over the short or medium term~\cite{economides_mobile_2017}. Money stored over a longer period of time becomes savings, so this sequential pattern would also occur if users were maintaining e-money in their mobile money accounts as a form of savings. This is less common~\cite{jack_mobile_2011}. Informal, over-the-counter, person-to-person transfers can also create the in-out motif. Often called a direct deposit, this action avoids the p2p transaction step; the sender cashes in to the recipient's account, rather than their own, with the cooperation of (or at the behest of) the depositing agent~\cite{singh_over_2016}. 

Sequential transactions following the in-out motif might also arise from opportunism on the part of mobile money agents. Encouraging and exploiting over-the-counter transfers is one of several ways by which agents could game mobile money systems so as to raise their earnings. More directly, agents can manipulate official commissions by acting strategically. Gaming is possible because agents earn a commission for facilitating both cash-ins and cash-outs, while providers earn revenue from this activity only from transaction fees charged on the cash-outs. Furthermore, agent commissions have a tiered structure. Agents can take advantage by splitting larger cash-in transactions into several smaller ones nearer the tier, effectively collecting multiple commissions for a single deposit. Since deposits incur no provider-imposed fee, this can be taken to an extreme and agents have been known to control user accounts for the sole purpose of earning themselves commissions. Under the commission structure of this particular provider, such brazen gaming would entail making many small deposits (maximizing the commission) and fewer large withdrawals (minimizing the provider-imposed fee). Since commission gaming comes at the expense of the mobile money provider, this whole range of actions are generally considered fraudulent. 

\section*{Methods}

Our approach is to analyze sequences of transactions. To do this, we define a data transformation that recovers empirical transaction sequences from financial transaction data. This ``follow-the-money'' transformation traces e-money from when it enters the mobile money system to when it leaves, noting the accounts that those funds pass through along the way. The result is a set of data objects that we call \textit{balance-respecting trajectories}, which represent a specific amount of money observed to move through a specific sequence of accounts via a particular sequence of transactions. 

\begin{figure}[h]
  \centering
  \includegraphics[width=1.0\textwidth]{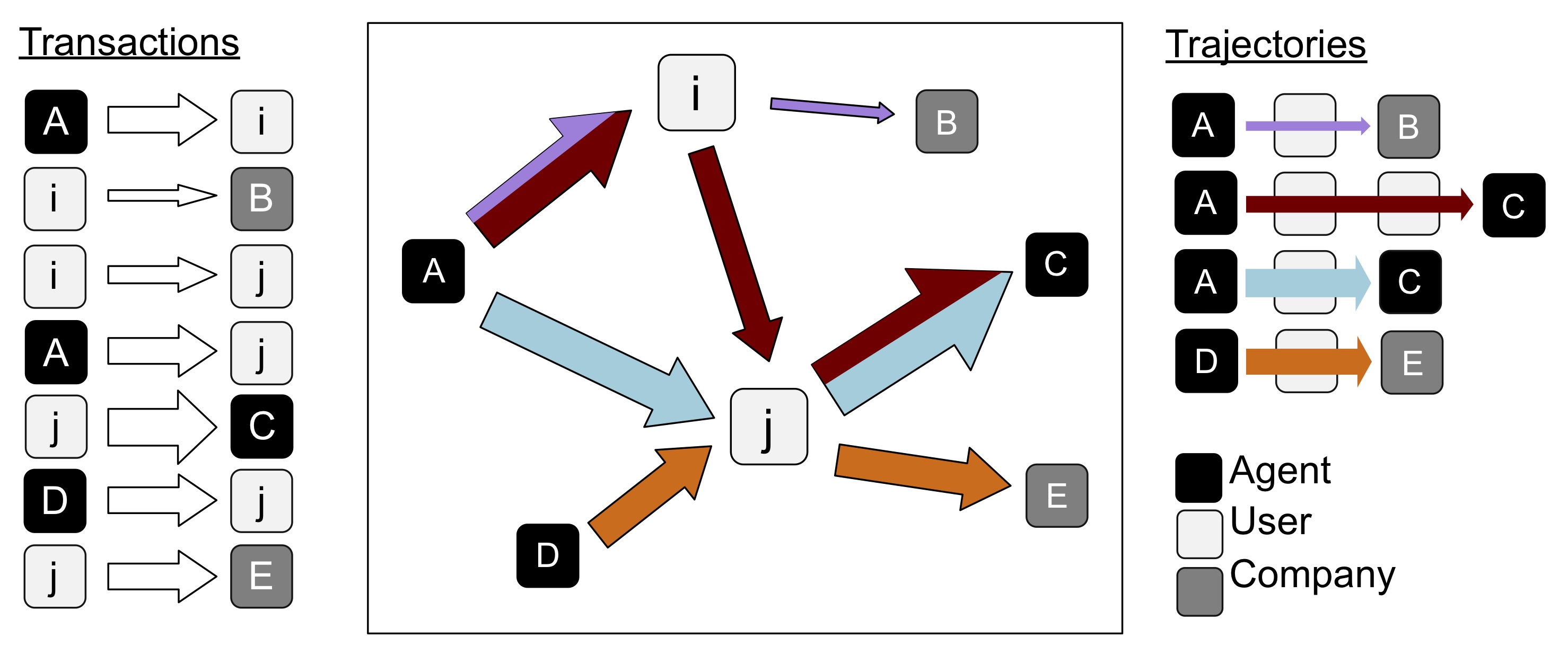}
  \caption{\textbf{Visual schematic of the ``follow-the-money'' transformation} An ordered series of deposits, transfers, payments, and withdrawals (left) create a transaction network (center). Arrows show the movement of e-money. Pieces of these transactions are allocated into balance-respecting trajectories (right). Trajectories represent the movement of e-money from depositing agents, through users, and on to companies and withdrawing agents. The purple, dark red, light blue, and orange trajectories follow micro-payment, digital transfer, in-out, and bill payment motifs, respectively.}
  \label{fig:transformation}
\end{figure}

Figure \ref{fig:transformation} illustrates this transformation for a simple series of transactions among mobile money agents, users, and companies. The arrows represent the movement of e-money, and the resulting set of balance-respecting trajectories follow motifs that correspond to those typical of mobile money systems. The orange trajectory follows the payment motif, where a user receives e-money via a cash deposit and subsequently uses it for a bill payment. A similar sequence where the funds are used for a mobile airtime or data purchase would form a micro-payment trajectory, shown in purple. The light blue trajectory follows the in-out motif, where a user makes a deposit only to cash the e-money back out without ever using it for a digital transaction.

\subsection*{Algorithmic implementation}

We implement our data transformation using a dynamic programming algorithm that funds outgoing transactions using e-money from prior incoming transactions. This algorithm records intermediate objects, \textit{branches}, that represent portions of transactions. \textit{Root branches} are portions of transactions that begin trajectories. As new transactions appear, existing \textit{branches} provide the funds to service them, building up a tree-like structure of references. \textit{Leaf branches} are portions of transactions that end trajectories. Once the algorithm reaches a \textit{leaf branch}, it has uncovered a complete trajectory and recursively traverses back toward the \textit{root branch}. The sequence of transactions, nodes, and durations encountered on the way back to the \textit{root branch} become the basic features of that balance-respecting trajectory. Follow-the-money is implemented concurrently for all accounts in the system as transactions appear in sequence or in continuous time. Please see the Supplementary Material for a deeper discussion of implementation decisions, details, and available alternatives. 

\textbf{Allocation heuristic} \hspace{12pt} Which existing funds to allocate to which outgoing transactions is not uniquely defined; money is fungible. In this paper, we allocate funds using a last-in-first-out or ``greedy'' heuristic. In the context of mobile money data, this heuristic ensures that a user who deposits \$100 through a mobile money agent, and then promptly pays a \$100 utility bill, will generate a straightforward \$100 e-money trajectory from the agent that processed their deposit, through their account, and on to the utility. 

\textbf{Network boundary} \hspace{12pt} What transactions are root, leaf, and regular branches will depend on the bookkeeping practices of the particular provider. We define the network boundary using the transaction types supplied in the data so as to trace all user-facing mobile money transactions. Transactions with a type that deposits e-money onto user accounts are defined to be \textit{root branches}, while payments and withdrawals are defined to be \textit{leaf branches}. The senders (recipients) of cash-in (cash-out) transactions are mobile money agents. The recipients of mobile airtime, mobile data, and bill payment transactions are corporations.

\subsection*{Accounting conserves money}

In building out balance-respecting trajectories we leverage a key feature of payment system records: financial transactions move money. Implicit in these data, then, are particular constraints that apply to the movement of money. Put simply, no one is allowed to spend the same dollar twice: if you were to use all your money to purchase a bike, then you would have none left with which to purchase a latte. Contrast this to a rumor or the flu, which you can absolutely share first with your bike mechanic and later with the barista. Moreover, it would be your own bank that steps in to decline your debit card at the coffee shop. Payment systems themselves are what enforce accounting constraints. 

Transactions that break accounting rules are not allowed, and payment systems see to it that they do not occur. In practice, accounting can be done in a decentralized manner (cash), a centralized manner (checking), or even algorithmically (blockchain). Either way, providers must enforce accounting or risk being forced to honor duplicated funds using money of their own. Because transactions can only be made using funds that already exist in the system, money is conserved. That accounting conserves money is even reflected in the terms we use to describe the dynamics of money, like \textit{flow} and \textit{circulation}. 

\subsection*{Precise mathematical representation}

On networks, conservative dynamics are represented as \textit{walk processeces}. These processes are studied in precise mathematical detail by researchers in network science~\cite{masuda_random_2017}. Our data transformation lets us represent financial transaction data from payment systems in existing mathematical terms: balance-respecting trajectories are \textit{observed instances of a weighted, continuous-time, node-centric, passive walk process on a temporal network}. While this particular type of walk process has yet to be studied, researchers have developed network analysis techniques using observed instances of simpler walk processes~\cite{pfitzner_betweenness_2013,rosvall_memory_2014}. As such, we can expect future methodological development to produce network analysis possibilities for balance-respecting trajectories that go well beyond those used in this paper.  

Masuda et. al. (2017) provide a taxonomy of random walks where a random transaction process would be a continuous-time passive random walk~\cite{masuda_random_2017}. The defining feature of these processes is that the temporal links of the network are what is moving the walkers; the transactions are the process. The time scale at which walkers are moving is the time scale at which the network itself changes, and this seriously complicates analysis. Many of the central results for random-walk processes on networks no longer hold when these two time scales are one and the same~\cite{perra_random_2012}. 

Masuda et. al. (2017) also distinguish between passive processes where activity centers on nodes, to those where activity centers on edges. They provide examples for \textit{edge-centric} passive walk processes, such as diffusion over temporal networks. Transaction processes are an example of the \textit{node-centric} variety; it is almost always either one counter-party or the other that initiates a transaction. Often it is the sender who initiates (ex. a payment), but there are transaction types where the recipient initiates (ex. a deposit). How we choose to represent nodes has a substantial impact on \textit{node-centric} walk processes. Indeed, different heuristics governing the movement of walkers through nodes can produce dramatically different walk statistics on temporal networks~\cite{saramaki_exploring_2015}. We can avoid this ambiguity with data directly from the walk process, itself. 

Click-streams, travel itineraries, and shipping logs are examples where data consists of known \textit{trajectories} that individually observed ``walkers'' followed through their network. Researchers have introduced methods for finding central nodes~\cite{pfitzner_betweenness_2013} and detecting communities~\cite{rosvall_memory_2014} on the networks revealed by such trajectory data. Taking each trajectory as a statistical observation, one can even create a higher-order network representation of the system that can approximate the observed trajectories using random walks~\cite{scholtes_higher-order_2016,lambiotte_understanding_2018}. 

In theory, one could use this approach directly on trajectories of individually marked bills through an economy~\cite{brockmann_scaling_2006}. In practice, however, individually marked bills are not exhaustively tracked and exhaustively tracked (ex. digital) money is not individually marked. What we can do is create balance-respecting trajectories. Transforming financial transaction data into this form is the first step towards using trajectory-based temporal network approaches to study the empirical structure of monetary flow. Adapting this full framework to trajectories with a meaningful size (amount of money) and duration (time spent in each account) is a promising topic for future work.  

\section*{Results}

We focus on the observed trajectories of e-money through the mobile money system that begin with cash-in transactions, and group them by the motifs they follow. Our first group combines all trajectories that end in a bill payment or micro-payment, and these motifs together capture 12.7\% of cash-in transactions. The next group encompasses the prototypical digital transfer motif, as well as similar motifs with more than one person-to-person transaction. These motifs also capture 12.7\% of cash-in transactions. Finally, we aggregate together all of the trajectories following the in-out motif, which reflects money storage or other activity that involves no digital transactions. 71.5\% of cash-in transactions follow this motif. 

For each of these groups of motifs, we create \textit{entry-exit networks} that describe the resulting movement of e-money through the system. The nodes in these networks are the mobile money agents or corporations at the start or end of each observed trajectory. The links between them are directed, and we give each deposit equal weight in calculating the aggregated link weight. The weighted out-degree of an agent corresponds to the number of cash-ins they facilitated that went on to follow a motif in that group. These networks represent the movement of money through the mobile money system as a whole, emphasizing the activity of users rather than the absolute flow of money, which would be strongly affected by the largest transactions. 

Please see the Supplementary Materials for finer details behind the analysis, visualization, and geographic inference.

\begin{figure}[htbp]
  \centering
  \includegraphics[width=1.0\textwidth]{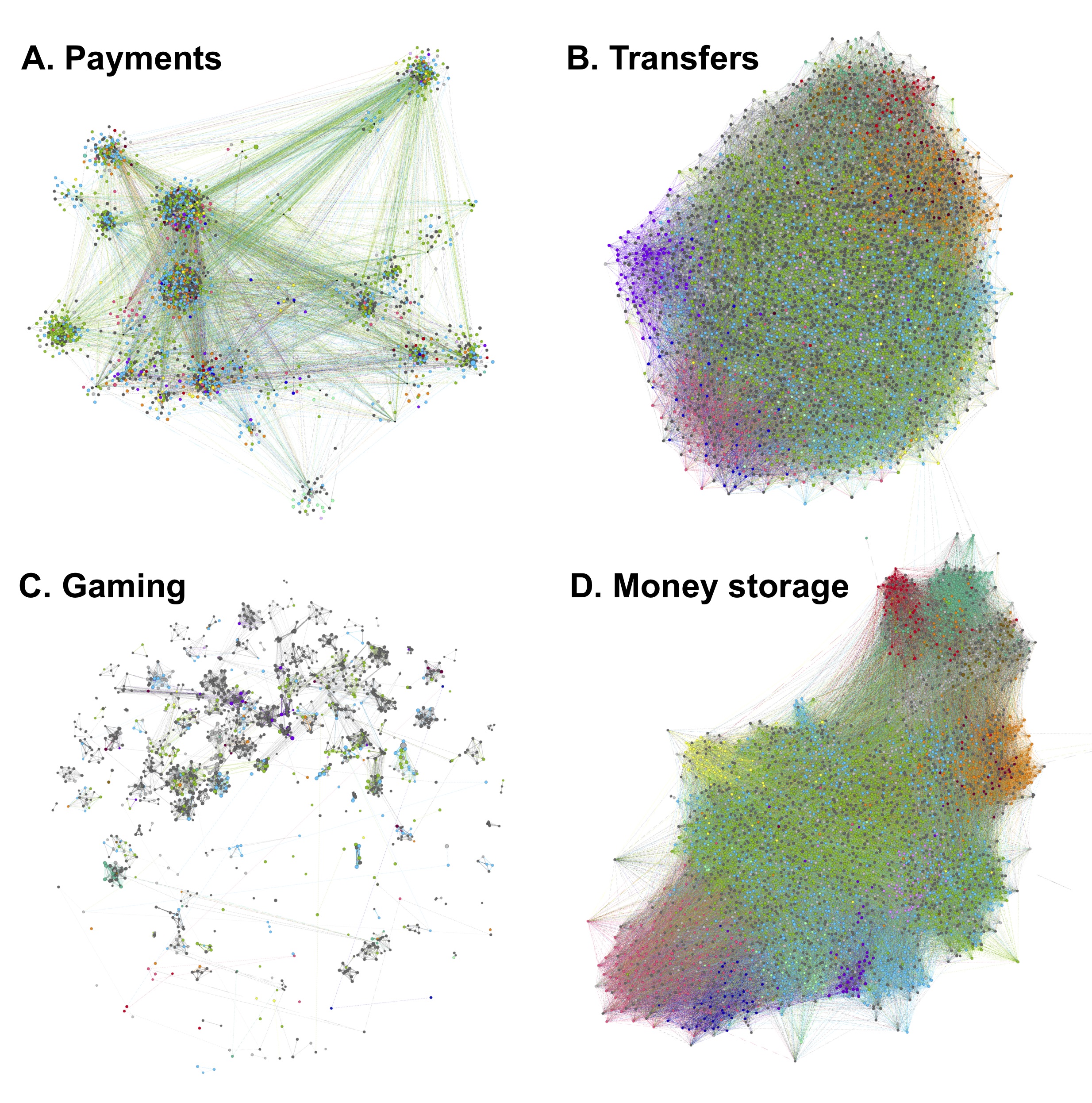}
  \caption{\textbf{Core structure of entry-exit networks} A visualization of the 4000-node core of the (\textbf{A}) digital payment and (\textbf{B}) digital transfer networks. The network of aggregated in-out trajectories shows a distinct structure among 1500 nodes of the innermost core likely engaging in (\textbf{C}) commission gaming and the 4000 nodes in the next tier facilitating (\textbf{D}) money storage or other non-digital activity. The top 10\% most significant links are displayed; isolates are hidden. Nodes are colored by geographic location at the highest sub-national administrative areas in the country, when known from cellular records. Agents who joined the network too recently for this (about half) are dark grey. Corporations are black points, and appear only at the center of the hubs in (\textbf{A}).}
  \label{fig:real_nets}
\end{figure}

\subsection*{Different economic ativities produce distinct network structures}

We find that mobile money facilitates four distinguishable economic actions: digital payments, digital transfers, commission gaming, and money storage or other non-digital activity. These activities move money through the system to form four decidedly different network structures: hub-and-spoke, amorphous, tightly grouped, and geographically assortative. Figure \ref{fig:real_nets} visualizes the weighted core of our three aggregated networks, showing the top 10\% most significant links within the core~\cite{eidsaa_s-core_2013,coscia_network_2017}. 

The network structure of (\textbf{A}) digital payments and (\textbf{B}) digital transfers appear strikingly different. When making digital payments, users move e-money from agents all over the country into the accounts of a handful of large corporations, who become the obvious hubs. In contrast, users making digital transfers move e-money through the system from everywhere to everywhere else in a manner that appears very close to random. Weighted k-core analysis reveals two distinct structural patterns within the in-out network. The innermost core of 1500 agents capture 13.1\% of all cash-in transactions as in-out motifs just among themselves (\textbf{C}). These agents form small and densely connected subgroups that are less connected to one another. This differs substantially from the structure among the next tier of 4000 agents that is indicative of the structure of the bulk of the in-out network and shows a general geographic assortativity (\textbf{D}).

\paragraph{Evidence for systematic commission gaming} We deem the innermost core of the in-out entry-exit network to reflect systematic commission gaming, predominantly, and proceed to consider it separately. These 1500 mobile money agents are a rather distinct set: only 7.6\% of them are also at the core of the payment or transfer networks, whereas this number is 51.7\% for the next 4000 by core number. The errant set includes less than 4\% of all agents, and they distinguish themselves with behavior that is consistent with engaging in systematic commission gaming. The average mobile money agent earns as commission 95.0\% of the fee revenue that they generate for the provider in facilitating cash-ins and cash-outs. The provider's break-even point is clearly somewhere below 100\%. On average, these 1500 agents earn as commission fully 231.9\% of the revenue they generate for the provider. We see evidence these agents are splitting deposits to reach such high commissions. While they serve about as many unique customers as does the average agent, these agents facilitate many times more cash-in deposits that are many times smaller. Moreover, a cash-in with one of these agents is four times as likely to fall within \$1 of a tier in the commission structure as one with an average agent. This is clear evidence of gaming. Finally, these agents facilitate almost no digital transactions; 95.0\% of their cash-ins follow the in-out motif. This means they may also be encouraging and exploiting over-the-counter transfers to raise their earnings further. 

\paragraph{Money storage and other non-digital activity} Without the errant contingent of agents, the in-out entry-exit network reflects regular mobile money activity that involves no digital transactions. The established explanation for such activity is money storage, but in our case it likely includes also mobile savings and over-the-counter transfers to some extent. We do not endeavor to distinguish between these actions, and especially not the intent behind them, as doing so would require stronger assumptions or additional data. In other mobile money systems, such as those with designated over-the-counter or savings services, it may be possible to distinguish these actions. 

\subsection*{Network structure of economic activities}

Network measures can quantify structural differences in the patterns of money flow created by the four distinguishable actions. We use the information-theoretic measure employed in the community detection algorithm Infomap to quantify the extent of sub-network structure~\cite{ding_community_2014}. This measure gives the average number of bits needed to describe one step in an infinite random walk on the network, and the algorithm exploits sub-network structure to minimize that value. We compare the value of the measure under compression to that of the uncompressed network. Random networks cannot be compressed, remaining near 0\% compression, while a network with increasingly rich multilevel subgroup organization would approach 100\% compression. To quantify the extent of geographic assortativity  we calculate the generalized modularity using the geographic locations of agents at the highest sub-national administrative areas in the country, when this could be inferred from cellular records~\cite{fortunato_community_2010}. This measure compares the amount of money moving between nodes within the same module to that which would be expected at random, and can range from -1 to 1. A value of 0 corresponds to random expectation; a value of 1 corresponds to a network where money moves only between agents within the same geographic area.

\paragraph{Pronounced subgroups in the structure of commission gaming} We have established that these mobile money agents are acting strategically, in that their behavior reflects the fee and commission scheme. These also move money amongst each other primarily within small subgroups, a curious network structure that may reflect deliberate coordination. This activity captures 13.1\% of cash-in transactions. Infomap achieves a full 63.2\% reduction in description length of the network, indicating that this network contains rich multilevel subgroup structure. Cash is often deposited and withdrawn from agents within the same groups of around ten agents. Although this is one particular case, this finding suggests that our analysis approach can surface particular kinds of strategic coordination, whether or not they are desirable, within payment systems. Future work on fraud detection in mobile money systems that flag individual in-out sequences with specific evidence of agent wrongdoing would allow researchers to further isolate and characterize commission gaming. 

\paragraph{Geographic assortativity in the structure of money storage and other non-digital activity} Regular in-out activity captures a remarkable 58.4\% of cash-in transactions, with an internal community structure driven by geography. This is an unexpectedly large share of all activity; the system is intended for digital transfers and they are the most popular service according to surveys~\cite{intermedia_financial_2016}. However, behavioral trace data carries different observational implications than do stratified surveys. In particular, the median mobile money transaction is not made by the median user, but rather by an especially active user who makes many transactions~\cite{wiel_implications_2012,mbiti_home_2013}. Money storage and other non-digital activity is prominent in the data because it reflects an important use-case among high-activity users. This activity is structured by geography. Infomap achieves a 5.6\% reduction in the description length as it finds community structure to the non-digital network. A generalized modularity of 0.27 by geographic location indicates that this structure is aligned with geography.  

\paragraph{Randomness in the structure of digital transfers} Digital transfer activity captures 12.7\% of cash-in transactions and forms an amorphous network with near-random structure. In contrast to the other use-cases, Infomap recovers next to no structure within the network of digital transfer activity. The algorithm achieves a negligible 0.06\% reduction in description length. Digital transfers show some geographic assortativity, with a modularity of 0.13, but little centralization. The 4000 agents at the core process 2.6\% of the cash-in transactions moving over it, which is not much more than naive expectation. That the structure of digital transfers is stubbornly amorphous is quite surprising, especially since mobile money has been unevenly adopted following existing contours of socioeconomic inequality~\cite{gsma_mobile_money_state_2015}. Much as those supporting the development of mobile money would like to pinpoint areas where digital circulation is succeeding especially well, this is not possible in this particular case. 

\paragraph{Prominent hubs in the structure of digital payments} Digital payment activity captures 12.7\% of cash-in transactions, and these funds end up paid to just over 300 corporate accounts. The large corporations who receive payments are ``hubs'' that hold prominent positions with respect to the movement of money within this mobile money system. The provider itself is one of these, as they are the recipient of e-money used to purchase mobile airtime and mobile data. 

\subsection*{Temporal structure of economic activities}

\begin{figure}[htb]
  \centering
  \includegraphics[width=0.9\textwidth]{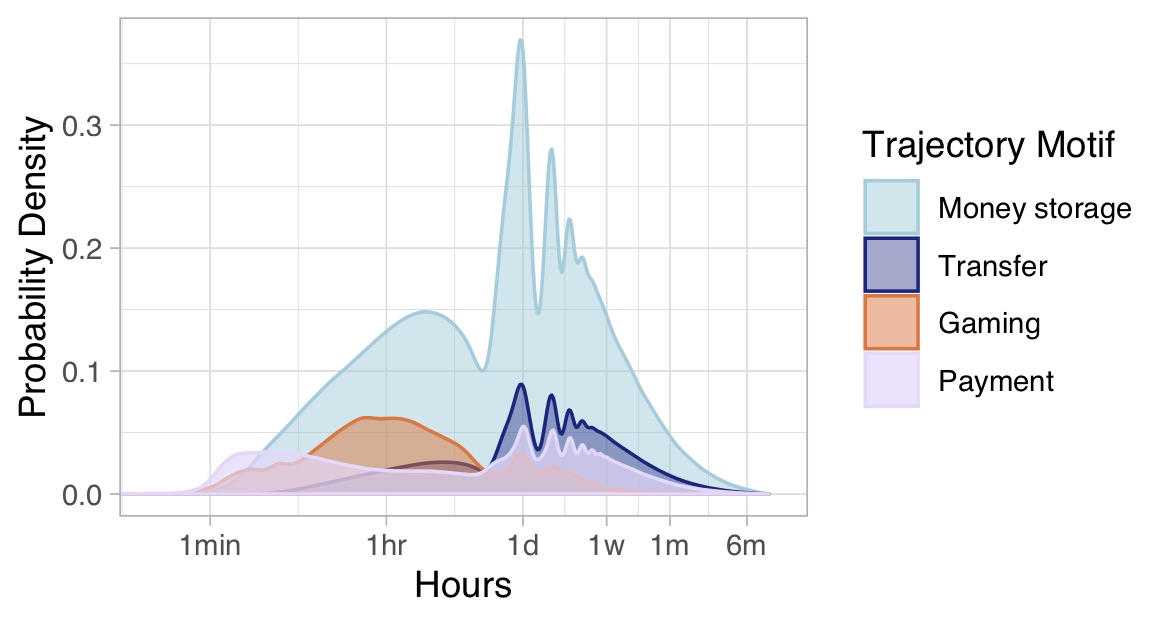}
  \caption{\textbf{Scaled distribution of trajectory durations} The distribution over the duration of time between cash-in transactions (that begin trajectories) and payment or withdrawal transactions (that end trajectories). Shown in color are the four distinguishable economic actions identified above. The distribution is weighted such that each cash-in contributes one observation, and the areas are scaled to reflect the proportion of cash-in transactions captured by each activity. The x-axes are log scaled.}
  \label{fig:dist_durations}
\end{figure}

We find that digital payments, digital transfers, commission gaming, and money storage or other non-digital activity also show different temporal structure. The trajectories corresponding to these activities move through the mobile money system over a period of time, and the profile of these durations differs substantially. Figure \ref{fig:dist_durations} shows the distribution of trajectory durations, scaled and weighted to reflect the proportion of cash-in transactions captured by each activity. 

The temporal structure of commission gaming and of digital transfers have the least overlap. Most commission gaming occurs within the same day while the majority of digital transfers take more than one day to move through the system. Digital payments show a bi-modal distribution, reflecting differences in two of the constituent actions: bill payments and micro payments. Cash-in deposits intended for bill payments routinely exit the system within a few minutes to an hour. Mobile airtime and mobile data purchases, on the other hand, are often made using the small sums that have remained in a mobile money account for days or even weeks. Money storage and other non-digital activity shows a very broad distribution, underscoring the difficulty in distinguishing actions that leave similar behavioral traces in the data.

Notably, we see wide variation in the duration distribution \textit{within} each activity. We know to expect differences \textit{across} use-cases in the amount of time e-money remains in the mobile money system. Mbiti and Weil (2013) estimate the turnover rate, or ``transactions velocity'', of mobile money in the M-Pesa system. They note that their estimate reflects an average over a hybrid system where money is both transacted rapidly and stored for longer periods of time.~\cite{mbiti_home_2013} They highlight a counter-intuitive observational effect: most of the e-money we see is used by those with rapid turn-over, but at any given moment most of the e-money in the system is held by those with slow turn-over. Our results show that we must contend with such effects also within any particular economic activity on mobile money systems. Indeed, the underlying duration distribution is logarithmic.

When values range across several orders of magnitude, the average becomes uncharacteristic of the distribution. The velocity of money is a theoretical concept defined by macroeconomic accounting relationships between money supply and price level, and is often treated as a single average value across an economy. It is related to the ``transactions velocity'', and there are methods that estimate the economy-wide velocity of money when average turnover rates differ across payment systems or sectors.~\cite{wiel_implications_2012,spindt_money_1985,leontief_money-flow_1993} It may be possible to extend these methods to incorporate heterogeneity also \textit{within} payment systems. Producing empirical measures comparable to the velocity of money at a sub-network scale, or even for individual accounts, is a promising direction for future research. 

\section*{Discussion}

In this paper, we find clear differences in the network and temporal structure of the movement of money across several distinct uses of a mobile money system. Several common use cases for mobile money---making a payment, transferring money, and storing funds---are interpretable as sequential combinations of mobile money transactions. Tracing funds through the system let us tease them apart. We group balance-respecting trajectories by the motifs they follow, relating observed individual-scale activity to system-scale network structure and back again. The resulting networks contain prominent hubs, random structure, geographic assortativity, and evidence of strategic behavior. Money moves through these patterns at highly heterogeneous rates. 

Our results give a hint as to what structures one would expect to find within the movement of money through an economy as a whole. The large corporations who receive payments are ``hubs'' in the mobile money network. We can expect systemically important companies to hold such prominent positions with respect to the movement of money within any economy. At the same time, geographic constraints on the opportunities for firms to do business are very real. Similarly, some amount of peer-to-peer activity that bypasses more centralized economic structures is to be expected almost anywhere. Strategic coordination is a field of its own within economics, and applying game theory to network formation can predict the existence of particular structural features.~\cite{jackson_social_2012}

Although the structural features we find are general, the particular activities we see reflect the affordances of mobile money, the incentive structure of this particular provider, and the economy of the country in which it operates. It is worth considering what appears to be largely missing from this mobile money system. An established bank in a more digitized economy might capture a wider range of economic activities such as receiving wages, buying products, paying suppliers, and servicing loans. A large-value payment system used by banks and major firms might capture investment decisions and financial trading. Those used by government agencies could capture taxation, allocation, and redistribution. The structure of money flow that results from any of these activities, and their relative share within a particular economy, are open empirical questions.

It will not always be possible to isolate different user actions as cleanly as done here. We are fortunate that the most common sequential transaction combinations in the data correspond directly to only one, or a few, well-documented use cases of mobile money. Furthermore, these are described in a robust substantive literature, technical publications, and available survey data.~\cite{economides_mobile_2017,gsma_mobile_money_state_2015,stuart_cash_2011,mbiti_home_2013,singh_over_2016,intermedia_financial_2016,jack_mobile_2011,wiel_implications_2012} On the other hand, other payment systems may have more detailed account labels or transaction descriptions. In some cases it may be possible to conduct surveys that directly ask about the intent behind common motifs. 

Even without any substantive information, the tools of network science would be useful for describing and interpreting the movement of money within any payment system. Network analysis can identify important nodes, pronounced subgroups, and community structure. These tasks are all intensely studied in network science, and could provide new ways to measure the economic power of large firms and the economic independence of regions from the flow of money. Ongoing advances in the field, particularly in trajectory-based approaches to temporal network analysis~\cite{scholtes_higher-order_2016,lambiotte_understanding_2018}, stand poised to expand the possibilities even further. 

Our methodology is also brings ``big data'' into reach for novel questions in monetary economics, inviting empirical research and theoretical development. We observe units of money as they move between accounts within a particular payment system, revealing the time dimension of money at the same level of granularity at which transactions occur. Turnover rates in this mobile money system differ widely even across instances of the same use-case. This underlying heterogeneity complicates estimation of the related theoretical concept, the velocity of money. This is defined via an accounting identity in standard monetary economics and is often assumed to have a single value across an economy. With expanded empirical tools, it may be possible to extend existing ways of incorporating differences in average velocity between payment systems~\cite{spindt_money_1985} and sectors~\cite{leontief_money-flow_1993} down to sub-networks, communities, or even individuals. 

In conclusion, this work can inform how macroeconomic models incorporate money. That an economy operates entirely within a single, universal payment system is often an implicit~\cite{mehrling_new_2010} or explicit~\cite{caiani_agent_2016} modeling assumption. From a practical standpoint, the actual financial system is more like a system of interacting payment systems that has yet to be mapped. Within each one, users will have highly heterogeneous rates of turnover which could introduce fragility in unexpected ways. From a theoretical standpoint, viewing the economy through the lens of \textit{the records that a universal payment system would collect} may be quite powerful in that economies as a whole could be represented as financial transaction networks. Links are transactions, and nodes are economic entities. Taking this perspective, the universal accounting logic that would apply highlights a deep similarity among everything from households to firms and government agencies. All economic entities must bring in more money than they spend in order to continue participating in the economy. How they do so, and how their efforts come together to create the economy, as a whole, becomes a compelling question worthy of both empirical and theoretical study. 

\printbibliography[title={References}]

\textbf{Acknowledgements:} The author acknowledges an extensive and fruitful collaboration with Guy Stuart of Microfinance Opportunities. She thanks Soren Heitmann, Shafique Jamal, John Irungu Ngahu, and Morne Van Der Westhuizen for comments over the course of this collaboration. The author acknowledges the International Finance Corporation (IFC) and the Partnership for Financial Inclusion for support and data access, under the direction of Soren Heitmann. For data preparation, the author thanks Cignifi Inc. Thank you to Geoff Canright and Brennan Klein for early discussions of this work, and to Alessandro Vespignani, David Lazer, Claudia Sahm, Bilge Erten, and Vincent Zountebier for key comments on the manuscript. \textbf{Funding:} This material is based upon work supported by the National Science Foundation Graduate Research Fellowship Program under Grant No. 1451070. Any opinions, findings, and conclusions or recommendations expressed in this material are those of the author and do not necessarily reflect the views of the National Science Foundation. \textbf{Competing interests:} Northeastern University holds US Provisional Patent 62/809,359 covering the data transformation process described in this work, from which the author may benefit financially. \textbf{Data and materials availability:} The software underlying this work can be found on Github, permission to use and modify is available from the author upon reasonable request. Data are available from the International Finance Corporation (IFC) based on the confidentiality agreement of the Partnership for Financial Inclusion with its clients. Restrictions apply to the availability of these data, which are not public and were used under license for the current study. Replication materials are available from the author upon reasonable request, subject to approval by the IFC. 

\vfill
\pagebreak

\section*{Supplementary Material} 

\subsection*{Mobile money data}

The data underlying this work consists of administrative transaction records collected in real time by a mobile money provider in East Africa. The data collection period ran from Jun 1, 2016 to Apr 1, 2017. The files were extracted by the provider, prepared and anonymized by Cignifi Inc., and provided to the author in their role as a consultant with Microfinance Opportunities by the International Finance Corporation under the Partnership for Financial Inclusion. Use of this data for the present study was ruled Exempt, Category \#4 by Northeastern University IRB\# 18-07-16.

Our analysis focuses on the most common activities. Less popular services are not shown in Table \ref{table:txn_types}, so the percentages do not add up to 100\%.

\paragraph{Geographic assignment} The assignment of agents to sub-national administrative areas of the country is based on inferences from mobile calling records. Transaction and calling records are linked via a shared unique identifier, a hashed phone number. The cellular calling records come from a period of six months that ended thirteen months before this data was collected, and includes the cellular tower through which outgoing calls were routed. Crucially, the provider shared a file that included the geographic location of most of the towers in this older data. Roughly half of the agents in the dataset appear in the earlier cellular records, and we assume they did not move in the meantime. This is not unreasonable, as agent often operate their business in conjunction with a retail shop or other fixed locations. We assume that the agents who do not appear in the earlier data joined the provider as agents within the year prior. Accounts were linked to the cellular tower through which a plurality of its outgoing cellular calls were routed over the full six months. We placed the GPS locations of these cell towers within administrative areas using QGIS and the shape files for the highest sub-national administrative areas of the country available from GADM, the Database of Global Administrative Areas~\cite{gadm_database_of_global_administrative_areas_global_2012}. 

\subsection*{Balance-respecting trajectories}
For this study, we trace the funds of all mobile money transactions with a transaction type deemed to be user-facing. We define the network boundary from the transaction types recorded in the dataset itself. Cash-in transactions, bulk payments from corporations to users, and deposits from ordinary bank accounts form one side of the network boundary: beginning trajectories. Cash-out transactions, ATM withdrawals, bill payments, micro payments, and withdrawals to ordinary bank accounts form the other side: ending trajectories. Person-to-person transfers and transactions involving merchants occur within the network boundary. Provider-facing transaction types are ignored unless they end existing trajectories. We do not consider corporations to be ``users'' of the system, since the transactions they recieve are subsequently handled on an ad-hoc basis by the provider. 

In this implementation, we allocate funds using a last-in-first-out or ``greedy'' heuristic. In the context of mobile money data, such allocation ensures that a user who deposits \$100 through a mobile money agent, and then promptly pays a \$100 utility bill, will generate a straightforward \$100 e-money trajectory from the agent that processed their deposit, through their account, and on to the utility. We also account for transaction fees charged by the mobile money provider and reference account balance information provided in the transaction data. We use a size cutoff at one unit of the local currency. We do not use a time cutoff. 

Our analysis focuses on the trajectories that begin as cash-in transactions. Table \ref{table:flow_types} shows a detailed breakdown of the transformed data by eight of the most common motifs. Not shown are more niche actions or incomplete trajectories, i.e. money that remains in the system at the end of the finite data collection window. For this reason, the percentages do not add up to 100\%.

\begin{table}[ht]
    \small
    \centering
    \begin{tabular}{|l|l|l|r|r|r|r|}
    \hline
    \textbf{Motif} & \textbf{Exit} & \textbf{Schematic}  & \textbf{Trajectories} & \textbf{Deposits}  & \multicolumn{2}{l|}{\textbf{Duration}} \\
    \textbf{} & \textbf{} & \textbf{}  & \textbf{} & \textbf{}  & \textbf{Median} & \textbf{Average} \\ \hline
    In-out   & Cash-out     & \includegraphics[scale=0.35]{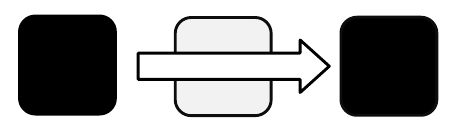} & 30.7\% & 71.5\%   & 10.1 hrs &  82.5 hrs \\ \hline
    Transfer & Cash-out     &\includegraphics[scale=0.35]{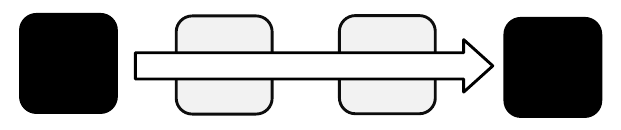} & 8.2\% &  11.3\%  & 27.8 hrs & 126.8 hrs \\ \hline
    Circulation & Cash-out     & \includegraphics[scale=0.35]{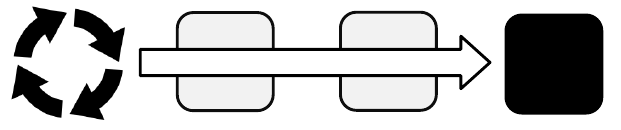} & 1.9\% &  1.4\%  & 90.3 hrs & 231.7 hrs \\ \hline
    Payment & Billpay         & \includegraphics[scale=0.35]{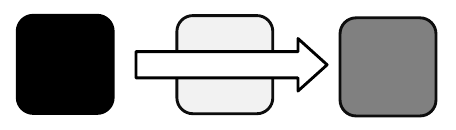} & 4.0\% & 6.6\% & 0.26 hrs & 27.1 hrs \\ \hline
    \shortstack[l]{Circulating\\payment} & Billpay         & \includegraphics[scale=0.35]{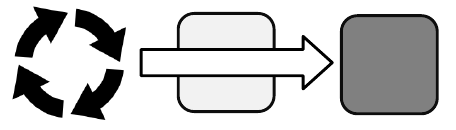} & 0.7\% & 0.3\% & 47.1 hrs & 152.0 hrs \\ \hline
    Micro-payment  & Topup/Data & \includegraphics[scale=0.35]{figures/payment.pdf} & 41.0\% &  5.1\%     & 34.7 hrs & 135.5 hrs \\ \hline
    \shortstack[l]{Circulating\\micro-payment}  & Topup/Data & \includegraphics[scale=0.35]{figures/payment-circ.pdf} & 13.1\% & 0.8\% & 79.8 hrs & 219.2 hrs \\ \hline
    \end{tabular}
    \caption{\textbf{Detailed summary of trajectory motifs} A summary of the trajectories observed when following all cash-ins through the mobile money system using the greedy (last-in-first-out) heuristic. Shown are eight of the possible motifs, where the arrows correspond to the movement of e-money. The number of trajectories observed to follow a motif is reported as a percentage of all unique trajectories beginning with a cash-in. The number of deposits is reported as the percentage of all cash-ins that went on to follow that motif; the median duration is the time in which half of those cash-ins had moved through that motif, and the average duration is the average across those cash-ins.}
    \label{table:flow_types}
\end{table}

\begin{figure}[ht]
  \centering
  \includegraphics[width=0.6\textwidth]{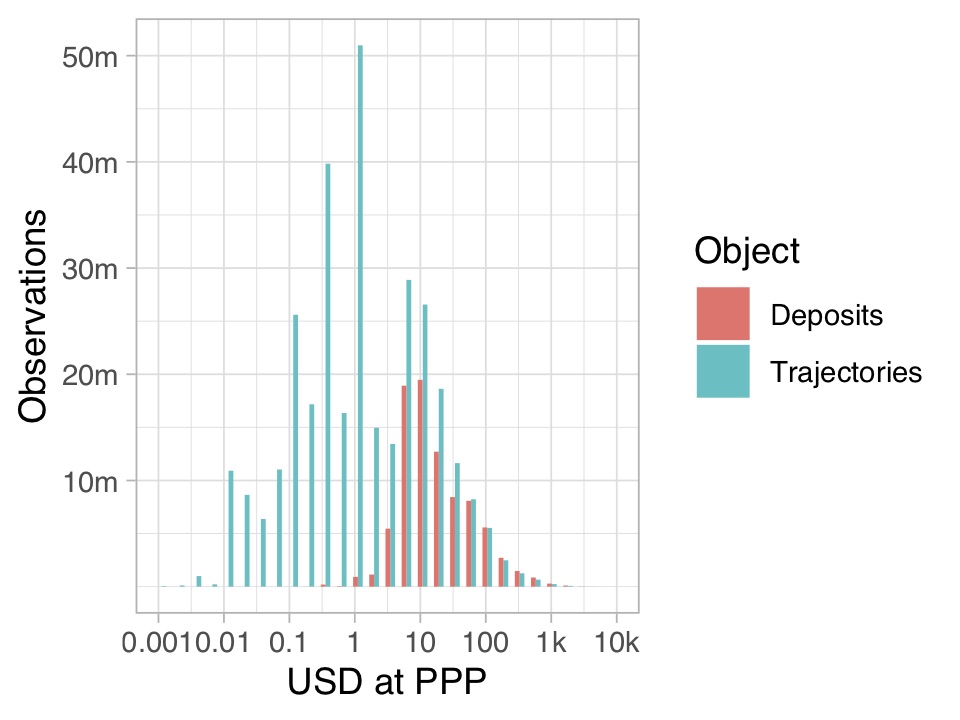}
  \caption{\textbf{Deposit vs. trajectory size} The histogram of trajectory sizes is compared to that of deposit transactions into the mobile money system. Both distributions cover several orders of magnitude and show a pronounced preference for round numbers. The x-axis is log scaled.}
  \label{fig:flows}
\end{figure}

Trajectories are interesting objects in their own right, and show pronounced heterogeneity in size. Deposit transactions into this system, shown in red in Figure \ref{fig:flows}, are \textit{already} distributed over several orders of magnitude -- there are deposits of around 1 USD at PPP and of several thousand. The process of tracing money through a payment system also creates more, smaller trajectories. In this particular system, we see more deposits of \$100 (PPP) or more than we do trajectories of that size, as downstream transactions split these deposits into many smaller trajectories. Particularly prominent in this data are the large number of small topup payments that split off many millions of small trajectories. 

\subsection*{Entry-exit networks}
We produce the aggregated entry-exit network using the eight motifs in Table \ref{table:flow_types}. The first encompasses all ``payment'' and ``micro-payment'' trajectories along with their circulating counterparts. The second includes all trajectories following a ``transfer'' or ``circulation'' motif. The last combines all trajectories following the ``in-out'' motif.

In building the \textit{entry-exit network} we aggregate together all balance-respecting trajectories with the same start- and end-points. These are mobile money agents or corporations (payment recipients). Self loops are allowed. For the link weight, we use the sum over the size of trajectories as a fraction of the initial cash-in transaction. This emphasizes the user activity involved in moving money, rather than reflecting mostly the largest transactions. A node's out-strength corresponds to the number of cash-in deposits that an agent \textit{initiated} that went on to follow a motif in the given group. The statistical techniques used to filter links~\cite{coscia_network_2017} and compress the network~\cite{ding_community_2014} both take link weights to be statistical ``observations''; cash-in deposits each contribute a total link weight of one and thus correspond to a single observation. 

It is worth noting that an analysis of other questions, such as cash re-balancing needs or profitability, would do better to use absolute amounts as link weights. Entry-exit networks aggregated using the absolute size of trajectories would represent the total flow of e-money. This would effectively re-scale link weights by the size of cash-in transactions, giving proportionately more weight to larger deposits. 

\paragraph{Network visualization} We use the weighted version of k-core, s-core~\cite{eidsaa_s-core_2013}, to isolate the core 4000 nodes for the ``payment'' and ``transfer'' network and the core 5500 nodes for the ``in-out'' network. We identify the the top 10\% of links within these cores according to ``noise corrected backboning''~\cite{coscia_network_2017}. Within the in-out core, the 1500 nodes with the highest s-core values have qualitatively different network structure and are shown separately. We use the open source graph visualization software Gephi.~\cite{bastian_gephi:_2009} Not shown are isolates and links below the 10\% threshold. For the ``payment'' network and ``commission gaming'' subgraph we used the OpenOrd layout with default parameters. This highlights and separates the tightly clustered groups of nodes in these networks. For the ``transfer'' and ``money storage'' networks we used the Force Atlas 2 layout with ``scaling'' set to 25, and otherwise default parameters. The nodes are sized by out-strength within each sub-plot on a negative quadratic spline. The nodes are colored consistently across the sub-plots, by the highest administrative area of the country to which the account was assigned. 

\subsection*{Identifying commission manipulation}
To conclude that the innermost core of agents are acting strategically in manipulating the commissions they earn, we compare the highlighted sets of agents according to several relevant indicators. The average values for agents within all highlighted sets are reported in Table \ref{table:agent_sets}. The average group size in the commission manipulation sub-graph is 4.9; the average such agent is in a group of size 9.9. 

\begin{table}[ht]
    \small
    \centering
    \begin{tabular}{|l|r|r|r|r|r|r|r|r|}
    \cline{1-9}
    \textbf{Network} & \textbf{Coreness} & \textbf{Users} & \textbf{Cash-ins} & \textbf{Amount} & \textbf{In-out} & \textbf{Near Tier} & \textbf{Gain} & \textbf{Established} \\ \cline{1-9}
              &       All &  836 & 1802 & \$50.59 & 66.63\% & 18.01\% &  95.04\% & 48.34\% \\ \cline{1-9}
    payments  &    1-4000 & 2491 & 5380 & \$43.16 & 64.50\% & 17.22\% &  89.88\% & 75.71\% \\ \cline{1-9}
    transfers &    1-4000 & 1517 & 2932 & \$50.12 & 66.48\% & 14.95\% &  80.32\% & 66.25\% \\ \cline{1-9}
    in-out    &    1-1500 &  872 & 8867 & \$ 9.76 & 94.95\% & 76.34\% & 231.94\% & 33.93\% \\ \cline{1-9}
    in-out    & 1501-4000 & 1834 & 3659 & \$45.40 & 68.78\% & 16.49\% &  82.59\% & 69.65\% \\ \cline{1-9}
    \end{tabular}
    \caption{\textbf{Description of highlighted sets of agents} A comparison of highlighted sets of agents, described by their coreness rank within the specified entry-exit network. The average number of unique users, cash-ins, and average cash-in amount is calculated over the values for the agents in that set. In-out is the average percentage of cash-ins to those agents that went on to follow the ``in-out'' motif. Near Tier is the average percentage of cash-ins to an agent that fall at or within \$1 above the commission tier (USD at PPP). Gain is the average percentage of the revenue earned by the provider on the cash-ins and cash-outs at that agent that is captured by the agent as commission. The percentage of agents in this set that are ``established'' is inferred by how many of them also appeared in a companion dataset from just over a year earlier.}
    \label{table:agent_sets}
\end{table}

\subsection*{Quantitative network analysis}
We use the full entry-exit networks to calculate the quantitative measures, considering the sub-graph of the 1500 suspicious agents separately. We run stand-alone hierarchical Infomap~\cite{ding_community_2014} with unrecorded teleportation at a 15\% probability, where the destination of jumps are chosen in proportion to the account's out-strength. The information-theoretic measure used in this algorithm -- description length -- allows us to quantify the extent of sub-network structure. This measure gives the average number of bits needed to describe one step in an infinite random walk on the network, and the algorithm exploits sub-network structure to minimize that value. We compare the value of the measure under compression to that of the uncompressed network, for the best of four runs. Random networks cannot be compressed, remaining near 0\% compression, while a network with increasingly rich multilevel subgroup organization would approach 100\% compression. We also calculate the generalized modularity~\cite{fortunato_community_2010} using the geographic locations of agents at the highest sub-national administrative areas in the country. This measure compares the amount of money moving between nodes within the same module to that which would be expected at random, and can range from -1 to 1. A value of 0 corresponds to random expectation; a value of 1 corresponds to a network where money moves only between agents within the same geographic area. We report the value as calculated across the subset of agents for whom geographic location is known. 

In Table \ref{table:network_measures} we report these measures for the aggregated networks corresponding to the four distinguishable mobile money actions. In this table, we report also the Gini coefficient across agents of the number of deposits entering the system, and those same trajectories exiting the system. Note that the recipients of digital payments, i.e. corporations, do not have a well-defined location nor do they facilitate cash-ins. As such, we do not calculate modularity or run Infomap on this network. 

\begin{table}[ht]
    \small
    \centering
    \begin{tabular}{|l|r|r|r|r|r|r|r|r|r|r|}
    \cline{1-10}
    \textbf{Network} & \multicolumn{2}{l|}{\textbf{Deposits}} & \multicolumn{2}{l|}{\textbf{Modularity}} & \multicolumn{2}{l|}{\textbf{Agent Gini}} & \multicolumn{3}{l|}{\textbf{Description Length}} \\ 
              & \textbf{Full} & \textbf{Core} & \textbf{Full} & \textbf{Known} & \textbf{Entry} & \textbf{Exit} & \textbf{Initial} & \textbf{Compressed} & \textbf{Reduction} \\ \cline{1-10}
    Payments    & 12.7\% & 19.5\% &  --  &  --  & 0.56 & 1.00 &  --       &   --      &   --    \\ \cline{1-10}
    Transfers   & 12.7\% &  2.6\% & 0.05 & 0.13 & 0.57 & 0.55 & 14.7 bits & 14.7 bits &  0.06\% \\ \cline{1-10}
    Gaming      & 13.1\% &  100\% & 0.12 & 0.45 & 0.98 & 0.98 & 10.1 bits &  3.8 bits & 62.30\% \\ \cline{1-10}
    Money storage & 58.4\% &  4.9\% & 0.13 & 0.27 & 0.54 & 0.55 & 14.8 bits & 13.9 bits &  5.58\% \\ \cline{1-10}
    \end{tabular}
    \caption{\textbf{Quantitative network measures} A comparison of the aggregated entry-exit networks. The number of cash-in deposits captured by this network is reported as a percentage of all cash-in transactions. The percentage of this total captured by the core of this network is reported as such. Modularity quantifies the geographic assortativity of the network using assignments of agents to sub-national administrative areas in the country, which is known for around half of all agents. This metric is unitless and ranges from -1 to 1. The Gini coefficient across agents quantifies the inequality of where trajectories begin and end. This metric is unitless and and ranges from 0 to 1. The description length is an information theoretic measure that decreases as the Infomap algorithm exploits sub-network structure to compress the network.}
    \label{table:network_measures}
\end{table}

\section{Follow-the-money implementation considerations} 

\subsection*{Allocation heuristics}
Allocation heuristics define how accounts in the system keep track of the money passing through them, determining what existing funds get assigned to an outgoing transaction. This work presents two such heuristics with particularly strong theoretical foundations: a mixing heuristic and a greedy heuristic. Which heuristic is most appropriate will depend primarily on the intended use of the transformed data. For example, a greedy heuristic is helpful in exploring intentional user choices. On the other hand, general analyses that invoke the concept of a random walk would be better served by the mixing heuristic. Data problems will tend to make the mixing heuristic less appealing. Either way, \textit{balance respecting trajectories} will preserve key time and accounting constraints on the system as a whole and keep trajectories interpretable as the flow of money.

\paragraph{Greedy Heuristic}
The greedy tracking heuristic represents each account as a stack of money, and the funds added most recently are the first to be spent.  More specifically, money from incoming transactions is added to the receiving account’s stack, on top of any existing funds in the account. To fill outgoing transactions, money is removed from the top of the account’s stack. This heuristic is in many ways the simplest possible, and is quite intuitive. An account that receives a \$100 deposit and promptly pays rent will generate a straightforward \$100 trajectory from whatever processed their deposit, through their account, and on to their landlord. It also has the attractive property that the results for a series of transactions is not affected by past transactions, including the size of the accounts in question. The individual paying their rent creates the same \$100 trajectory irrespective of whether they have \$10 in their account or \$10,000. In some ways, this parallels how people may think about money and thus introduces a stylized representation of savings into the system. Colloquially, dusty old money collects at the bottom of an account until the user needs to dip into it. 

\paragraph{Mixing heuristic}
The mixing heuristic represents each account as a well-mixed pool. Under this formulation, money from incoming transactions joins existing funds in the account with no added distinction. To fill outgoing transactions, money is drawn evenly from the account’s pool. Each earlier incoming transaction contributes to outgoing ones in proportion to the total balance of the account. This heuristic has the attractive property that it recovers all possible paths that a unit of money could have taken through the system, with a weight corresponding to the path’s relative likelihood. This approaches the notion of a random walk in network science. It also dovetails nicely with the economic conception of money as perfectly fungible---any unit of currency is considered entirely equivalent to any other.  Caution is warranted, however, because this heuristic is not independent of the past. Only completely emptying an account makes certain paths through it impossible; accounts with small balances constrain the universe of possible paths to a greater extent. With a finite time window into the system we must know or infer the initial balance of every account for the mixing heuristic work as advertised. 

\begin{figure}[ht]
  \centering
  \fbox{\includegraphics[width=0.47\textwidth]{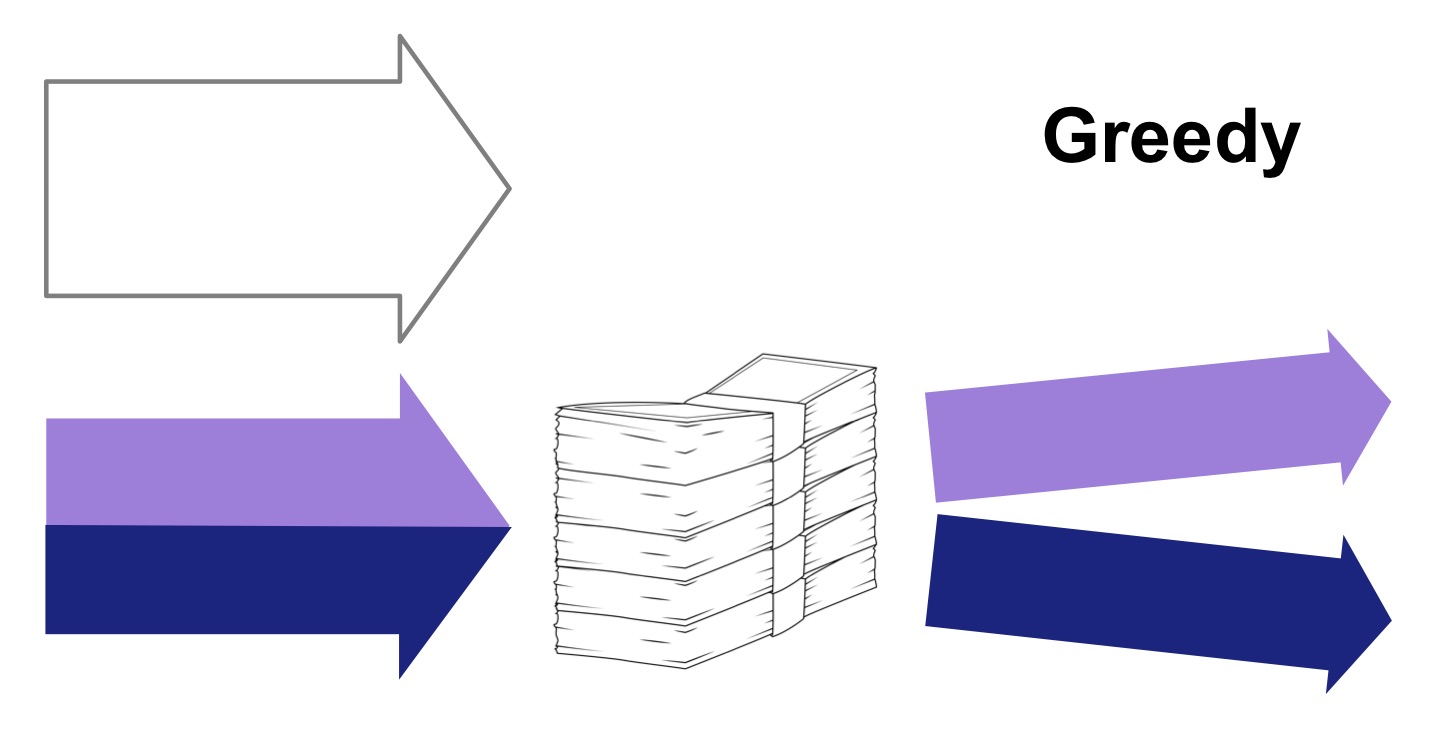}}
  \fbox{\includegraphics[width=0.47\textwidth]{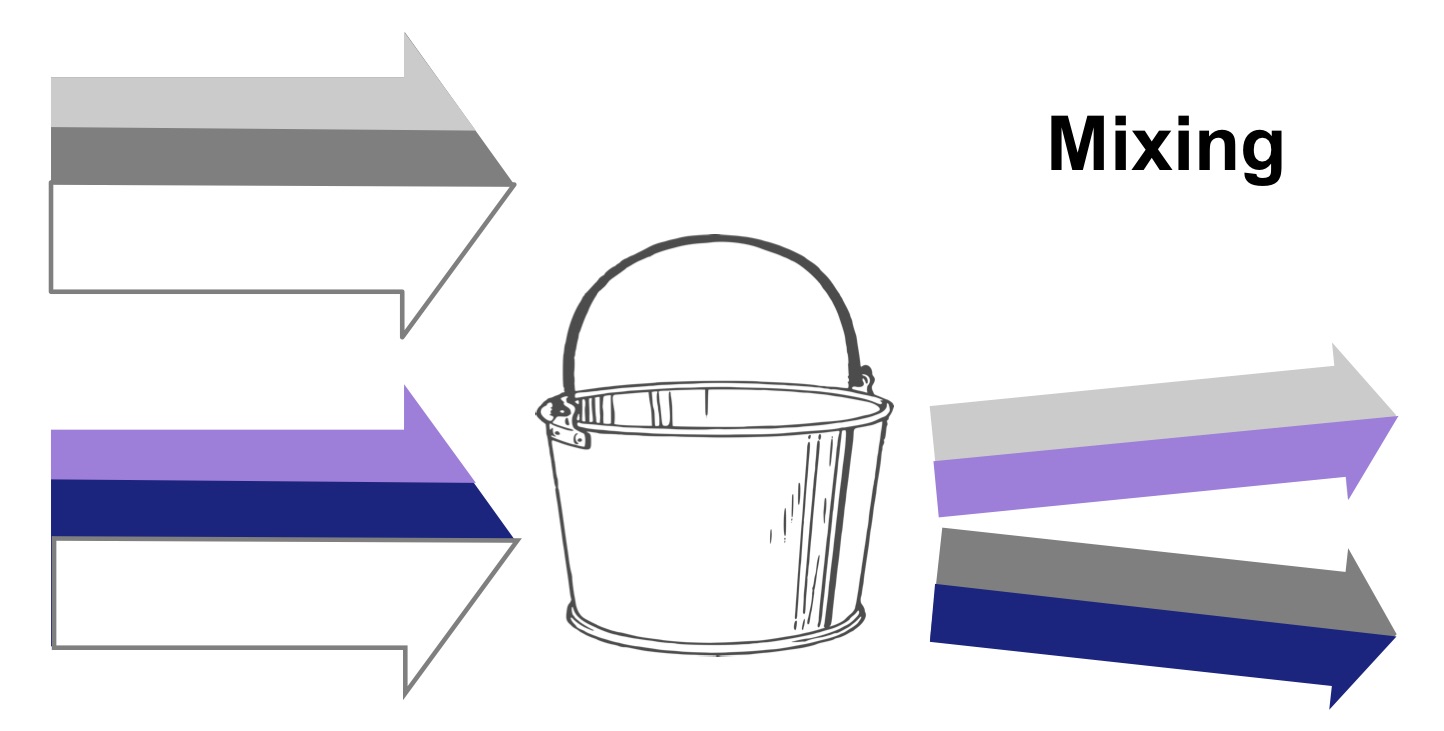}}
  \caption{\textbf{Allocation heuristic example} An illustration of the allocation outcomes for a simple series of transactions involving one account. This highlighted account is represented by a stack for greedy allocation, and a pool for well-mixed allocation. The account receives two \$100 transactions, and later sends \$50 to two different accounts. The greedy heuristic uses the most recent incoming transaction to fund the outgoing transactions, creating two \$50 trajectories. The mixing heuristic pulls evenly from both incoming transactions, creating four \$25 trajectories. Empty arrows are as-yet un-allocated funds.}
  \label{fig:twostep}
\end{figure}

\subsection*{Network boundary}
Payment systems are rarely fully contained. Many providers allow users of the system to deposit and withdraw from their individual accounts, leaving the total balance of the system to fluctuate with use. This means that most payment systems have a user-facing side where the movement of money is user-driven, and a provider-facing side that accommodates users' deposits and withdraws. The payment system domain already provides the terminology we need to describe the network boundary: ``deposits'' and ``withdrawals'' add or remove money from the system while ''transfers'' circulate money within it. Although bookkeeping practices vary, these distinctions are often salient for providers and thus feature prominently in transaction records. Providers generally incur much lower costs when users transfer money within their system than when users add or remove money from their system, which requires the provider to interface with other payment systems. For example, maintaining locations where users can obtain physical currency (branches, agents, ATMs, etc.) is costly, and other providers generally charge for the use of their systems (bank wires, SWIFT, etc.). 

Defining a system boundary allows the follow-the-money transformation to trace funds only for user-driven activity. As long as we have a way to categorize transactions as ``deposits'', ``withdrawals'', or ``transfers'' it is possible to define a boundary, and thus we can say where \textit{balance respecting trajectories} begin and end. Exactly where to delineate the system boundary will depend on the details of the provider’s systems and how to categorize transactions will depend on the idiosyncrasies of their bookkeeping practices. Of course, placing the boundary around user-facing accounts is not the only option and other choices will be appropriate for other analyses. 

The code presented alongside this paper presents options for defining the boundary of the system based on known transaction types, known account types, inferred account types, and several combinations. Not defining a boundary treats the system as fully contained.

\subsection*{Further nuances}
\paragraph{Transaction Fees}
Payment system providers often charge transaction fees, which are paid by users for access to the service. If each transaction contains information on the fee or fees that users pay to use the service (ie. the revenue the provider is generating from running the service), then money must be diverted to pay them. The size of trajectories will then decay as they move through the system, with more and more of it allocated to fees. This means that the size of the trajectory is also indexed by the step along the trajectory. In its implementation, this decay is incorporated as a list of revenues that are paid at that step of the trajectory. 

Note that providers may have different conventions for recording fee/revenue. One option is to include a column that specifies the fee, which is pulled from the sender's account alongside the transaction amount (ie. the sender pays). Another option is for the specified fee to be removed from the transaction amount before it enters the recipient's account (ie. the recipient pays). It is also possible for providers to charge both kinds of fees, or to note the fees they charge as entirely separate transactions. 

\paragraph{Balance information}
If the transaction file contains information on the balance of accounts at the time of a transaction, this can be useful. Discrepancies between the algorithm's internal accounting and the known balances can expose missing transactions. Although it is impossible to know where the money came from, it can be useful to note the discrepancy by inferring the existence of a deposit or withdrawal that brings the internally calculated balance back into line with what is given. Note that accounting imperatives of the transaction override even a given balance.

\paragraph{Size cutoff}
It is sometimes useful to limit the granularity at which money is followed. The mixing heuristic, in particular, will create many tiny trajectories rather quickly when many accounts maintain non-zero balances. A closed system would also end up with increasingly many, increasingly tiny, trajectories over time. This will eventually overwhelming memory capacity, and so it is useful to place a lower bound on the size of trajectories. To do so, anytime allocating funds to a transaction would create a branch that is too small the existing branch instead becomes a \textit{leaf branch}, ending the trajectory. 

\paragraph{Time cutoff}
Within the framework of \textit{balance respecting trajectories}, it is straightforward to introduce \textit{time-cutoffs}. With \textit{time-cutoffs} accounts are directed to forget the history of money that has remained in their account for longer than a that period of time. When such money is subsequently transacted, this account becomes the starting point of a new \textit{trajectory}. This would also be useful in cases where the fully resolved algorithm becomes computationally untenable. 

\subsection*{Memory usage and runtime}
The transformed data will almost certainly be larger than original. Follow-the-money does not keep the full data in memory, but does command considerable resources especially under the mixing heuristic. It cannot be easily parallelized. For the data set used in this work the greedy heuristic took ~12 hours and used ~20G memory while the mixing heuristic took ~48 hours and used ~60G of memory.

\end{document}